\documentclass[inactive,preprint,tightenlines,superscriptaddress, amsmath,amssymb,
 aps,prl,notitlepage]{test_revtex4.1}
\DeclareMathOperator*{\argmin}{arg\,min}
\usepackage{tablefootnote}
\usepackage{threeparttable}
\usepackage[normalem]{ulem}
\usepackage{graphicx}
\usepackage[utf8]{inputenc} % allow utf-8 input
\usepackage[T1]{fontenc}    % use 8-bit T1 fonts
\usepackage{hyperref}       % hyperlinks
\usepackage[capitalise]{cleveref}
\usepackage{url}            % simple URL typesetting
\usepackage{booktabs}       % professional-quality tables
\usepackage{amsfonts}       % blackboard math symbols
\usepackage{nicefrac}       % compact symbols for 1/2, etc.
\usepackage{microtype}      % microtypography
\usepackage{lipsum}
\usepackage{placeins}
\begin{document}

\title{Predicting Growth Rate from Gene Expression}

\author{Thomas P. Wytock}
\affiliation{%
 Department of Physics and Astronomy, Northwestern University, Evanston, IL 60208, USA
}%
% \altaffiliation[Also at ]{Department of Physics and Astronomy, Northwestern University, Evanston, IL 60208, USA}%Lines break automatically or can be forced with \\
\author{Adilson E. Motter}%
\affiliation{%
 Department of Physics and Astronomy, Northwestern University, Evanston, IL 60208, USA
}
\affiliation{
 Northwestern Institute on Complex Systems, Evanston, IL 60208, USA
}%
\affiliation{
Physical Sciences-Oncology Center, Northwestern University, Evanston, Illinois 60208, USA
}
%\author{Thomas P.~Wytock\setcounter{savecntr}{\value{footnote}}\thanks{Department of Physics and Astronomy, Northwestern University, Evanston, IL 60208} \\
%\texttt{t-wytock@northwestern.edu}
%  %% examples of more authors
%   \And
%Adilson E.~Motter\setcounter{restorecntr}{\value{footnote}}%
%  \setcounter{footnote}{\value{savecntr}}\footnotemark% Print footnotemark
%  \setcounter{footnote}{\value{restorecntr}} 
%   \ \thanks{Northwestern Institute on Complex Systems, Northwestern University, Evanston, IL 60208}
%   \ \thanks{Chicago Region Physical Sciences-Oncology Center, Northwestern University, Evanston, IL 60208}\\
%  \texttt{motter@northwestern.edu} \\
%}

%\authorcontributions{AEM and TPW conceived the research. 
%TPW retrieved data and designed the computational method.
%AEM and TPW wrote the manuscript. }
%\authordeclaration{The authors declare that no competing interest exists.}
%\correspondingauthor{\textsuperscript{2}To whom correspondence should be addressed. E-mail: motter@northwestern.edu}

%

%\dates{This manuscript was compiled on \today}
%\doi{\url{www.pnas.org/cgi/doi/10.1073/pnas.XXXXXXXXXX}}

%\verticaladjustment{-2pt}

%\thispagestyle{firststyle}
%\ifthenelse{\boolean{shortarticle}}{\ifthenelse{\boolean{singlecolumn}}{\abscontentformatted}{\abscontent}}{}

\begin{abstract}
Growth rate is one of the most important  and most complex  phenotypic characteristics of unicellular microorganisms, 
which determines the genetic mutations that dominate at the population level, and ultimately whether 
the population will survive. Translating changes at the genetic level to their growth rate consequences
remains a subject of intense interest, since such a mapping could rationally direct experiments to optimize
antibiotic efficacy or bioreactor productivity. In this paper, we directly map transcriptional profiles to growth rates 
by gathering published gene-expression data from \emph{Escherichia coli} and \emph{Saccharomyces cerevisiae} 
with corresponding growth-rate measurements. Using a machine-learning technique called $k$-nearest-neighbors 
regression, we build a model which predicts growth rate from gene expression. By exploiting the correlated nature 
of gene expression and sparsifying the model, we capture 81\%  of the variance in growth rate of the \emph{E. coli}
dataset while reducing the number of features from over 4,000 to nine.  In \emph{S. cerevisiae}, we account for 89\% of the variance in growth rate while reducing from over 5,500 dimensions to 18.  Such a model provides a basis for selecting successful strategies from
 among the combinatorial number of experimental possibilities when attempting to optimize complex phenotypic traits like growth rate. 
\vspace{12pt} \\ 
 Wytock, T. P., \& Motter, A. E. (2019). Predicting growth rate from gene expression. \emph{Proc. Natl. Acad. Sci. USA}, 116(2), 367--372
 \end{abstract}
\maketitle

\section*{Significance summary}
Connecting genetic changes to organismal function has been a central problem of biology for decades. Understanding the genetic underpinnings of functional traits like growth rate remains incomplete despite efforts to uncover metabolic and gene regulatory networks. Here, we leverage correlations derived from large-scale datasets of \emph{E. coli} and \emph{S. cerevisiae} to construct a mapping between gene  expression and growth using the $k$-nearest neighbors technique. Our mapping can predict growth rate more accurately than previous methods, while compressing gene-expression data from thousands of genes to tens of features without requiring network structure identification. This model can be applied to generate hypotheses, design experiments, and reduce the amount of trial and error in research.

\section*{Introduction}

%\vspace{-.2cm}
Mapping genotype to phenotype remains a central challenge in molecular biology. 
In the past two decades, complex networks have emerged as a tool to organize the vast amount of data generated by genomic technologies toward mapping biochemical patterns to whole-system function. Applications include the meta-analysis of genetic interactions across organisms to find biologically conserved structures~\cite{Ravasz2002}, the network analysis of gene annotation relationships to interpret the expression changes in gene sets~\cite{Glass2012,Ku2012}, the curation of genetic relationships into gene-regulatory networks to predict the outcome of proposed interventions~\cite{Zanudo2015, Cornelius2013}, and the reconstruction of metabolic networks on which metabolic capacity is calculated through constraint-based models (CBMs)~\cite{Edwards2000,Forster2003a,Segre2002}.  These tools have also been used to explain gene essentiality and epistasis~\cite{Forster2003b,Gerdes2003,Segre2005a}, and to study disease progression and treatment~\cite{Zanudo2015,Schlauch2017}. 

Though successful, these strategies all rely on the labor-intensive task of determining the network of relationships between genes. They resolve the network structure through a combination of  aggregation of prior knowledge and targeted experimentation, but the benefit that network models provide by structuring data are limited by the problems they address or the conditions under which they apply. For example, annotation methods are associative and qualitative, which limit the potential for causal attribution and inter-study comparison, respectively. Meanwhile, precise dynamic models of gene-regulatory networks often require the (challenging) measurement of \emph{in vivo} kinetic parameters or other condition-specific quantities to validate the dynamical rules. On the other hand, CBMs require experiments with well-defined media and measurements of metabolic uptake rates for flux-balance analysis (FBA) to yield accurate maximal rates of biomass production. In addition, FBA assumes that the cell directs its metabolic activities to maximize cell growth and is fully adapted to its environment both before and after a perturbation~\cite{Edwards2000,Fong2004}, although alternate methods have been developed that relax this restrictive requirement on the final state~\cite{Segre2002,Shlomi2005} or more generally~\cite{Mahadevan2003}.

In this paper, we establish a complementary method to predict growth rate using only gene-expression data, which we refer to as Model-Independent Prediction Of Growth Using Expression (MI-POGUE). Even though our method focuses on growth rate,
it provides a novel strategy to answer the more general questions of how whole-cell gene expression affects phenotypic changes
and thus of how to convert genome-wide observations into quantitative phenotypic predictions.  
The novelty and flexibility of MI-POGUE derive from using an effective model of genetic interactions in lieu of relying on prior knowledge or specialized experiments.

We develop MI-POGUE by retrieving large datasets of gene expression and growth rate in \emph{E. coli}~\cite{Carrera2014} and 
\emph{S. cerevisiae}~\cite{Hughes2000,Airoldi2009,Airoldi2016,Slavov2011,Lu2009,Kemmeren2014}. Comprising thousands of individual observations, these datasets allow the direct measurement of gene-gene correlations present in cells, 
which form the basis of an effective model of genetic regulation. In our approach, we transform the gene-expression data into weighted combinations of genes derived from the gene-gene correlations called  ``eigengenes''~\cite{Alter2000}, and predict growth rate by averaging the growth rates associated with the gene-expression profiles most similar to a given target profile---a technique known as $k$-nearest neighbors~(KNN) regression \cite{Altman1992}. The efficacy of MI-POGUE is substantiated by comparing it with state-of-the-art methods for predicting growth rate.

The data-driven conception of MI-POGUE sidesteps the network identification problem 
while still accounting for all observed changes to intracellular networks in response to perturbations. Given that nonlinearity allows small changes in part of the cell to effect large changes in another part, broad-based strategies like MI-POGUE that account for changes across the whole genome promise to open new lines of inquiry in investigating fundamental systems biology, as well as in engineering microorganisms,  and designing antibiotics. 

%\vspace{-.4cm}
\section*{Results}
\label{Results}
\subsection*{Dataset overview}
MI-POGUE requires a number of paired gene-expression and growth-rate measurements large enough to form a representative sample of potential organismal growth conditions to provide accurate growth estimates. 
 We apply MI-POGUE to both \emph{E. coli} and \emph{S. cerevisiae}.  Table~\ref{tab:datasettab} establishes that the size of the dataset we consider is unusually large for each organism.

\begin{table}[t]
\centering
\begin{threeparttable}[t]
    \caption{Overview of datasets}
    \begin{tabular}{p{5cm} p{3cm} p{3cm}}
\hline
Description & \emph{E. coli} & \emph{S. cerevisiae} \\ 
 \hline
Gene-expression profiles &  2,196   &  2,170 \\
Growth-rate measurements & 589 & {107} \\ 
\hline
\end{tabular}
\label{tab:datasettab}
\begin{tablenotes}
    \item[] \emph{E. coli} data from ref.~\cite{Carrera2014}.  \emph{S. cerevisiae} data from refs.~\cite{Hughes2000,Airoldi2009,Airoldi2016,Slavov2011,Lu2009,Kemmeren2014}.
\end{tablenotes}
\end{threeparttable}
%\vspace{-0.5cm}
\end{table}

The \emph{E. coli} dataset is derived from ref.~\cite{Carrera2014} and includes a broad sample of environmental conditions, measuring the effects of heat shock, hypoxia, or adaptive evolution on a variety of carbon sources in addition to over 150 genetic perturbations. The primary substrains of \emph{E. coli} K12 featured in these experiments are MG1655 and BW25113.  The \emph{S. cerevisiae} dataset, which serves to demonstrate its applicability to eukaryotes, comprises experiments performed in chemostats with various environmental stresses and nutrient limitations with gene-expression data taken from refs.~\cite{Hughes2000,Airoldi2009,Airoldi2016,Slavov2011,Lu2009,Kemmeren2014} and growth rate taken from refs.~\cite{Airoldi2009,Airoldi2016,Slavov2011,Lu2009}.
 The metadata annotating the experiments is curated from these references, as described in the Methods, and they are provided with MI-POGUE's source code~\cite{Wytock2018}.  %~\ref{data:source}.
%\vspace{-.2cm}
\subsection*{Eigengene estimation}

The expression between genes is highly correlated~\cite{Alter2000}, implying that each gene's expression
depends on its neighbors in the network.
Here, we derive eigengenes, which are combinations of genes that reorganize expression according to interdependencies 
implicitly mediated by the gene regulatory network.
Let $g^{i}$  and $\vec{v}^{\,i} = \{ v_j \}^{i}$ be the growth rate and gene expression, respectively, of the $i^{th}$ experiment, and let $j$ be an index over genes, and let $\vec{g} = \{ g^{i} \} \  i=1..E$ and $\vec{v} = \{ \vec{v}^{\,i} \} \  i=1..E$ be the set of all growth-rate measurements and their associated gene-expression profiles, respectively. Furthermore, let $E$ be the total number
of experiments including a growth-rate measurement, $N$ be the number of gene-expression measurements used to estimate correlations,
and $J$ be the total number of genes common to each measurement.

We estimate the correlations between genes based on all the available expression data 
(regardless of whether it had associated growth rate or not) and calculate the eigenvectors. Briefly,
we compute the gene-gene (Pearson) correlation matrix $\mathbf{C} = C_{jk}$ where $C_{jk}$ is the
Pearson correlation coefficient between the expression of the $j^{th}$ and $k^{th}$ genes.
The correlation matrix is square ($J$ by $J$) and symmetric under exchange of indices. We diagonalize the matrix, 
\begin{equation}
 \mathbf{C} = \mathbf{P}\mathbf{D}\mathbf{P}^{-1}, \label{simtrans}
\end{equation}
resulting in the matrix $\mathbf{P}=P_{lj}$ in which each column $\mathbf{P}^{l}$ corresponds to the $l^{th}$ eigenvector
while each row $\mathbf{P}_j$ reflects the $j^{th}$ gene's projection onto the set of eigenvectors. The 
diagonal matrix $\mathbf{D}=(D_{ll})$ indicates the amount of correlation occurring along the $l^{th}$ column of 
$\mathbf{P}$. Because the correlation matrix is symmetric, we have
\begin{equation}
\mathbf{P}^{-1}=\mathbf{P}^T. \label{transpose}
\end{equation} Any expression profile may be projected onto the correlation eigenvectors by matrix multiplication:
\begin{equation}
 \vec{\nu}^{\,i} = \mathbf{P}^T \vec{v}^{\,i}. \label{proj}
\end{equation}
We call the $\vec{\nu}^{\,i}$ \emph{eigengenes}, a portmanteau of ``eigen'' (proper) and ``gene''~\cite{Alter2000},
because they represent independent (that is, non-redundant) variations in gene-expression space.
Therefore, increasing or decreasing the magnitude of one eigengene's expression ($\nu_l^{i}$), leaves the other projections unchanged ($\nu_m^{i}\,, m \neq l$). 
In contrast, changing the expression of the $j^{th}$ gene, $v_j$, would result in changes in other genes, modulated by $\mathbf{P}_j$, allowing this change to have wide-ranging impacts across the gene-expression profile.
%\vspace{-.2cm}
\subsection*{Restriction of KNN models to the most informative eigengenes}

\begin{figure*}[t]
 %\vspace{-1cm}
\begin{center}
\includegraphics[width=\textwidth]{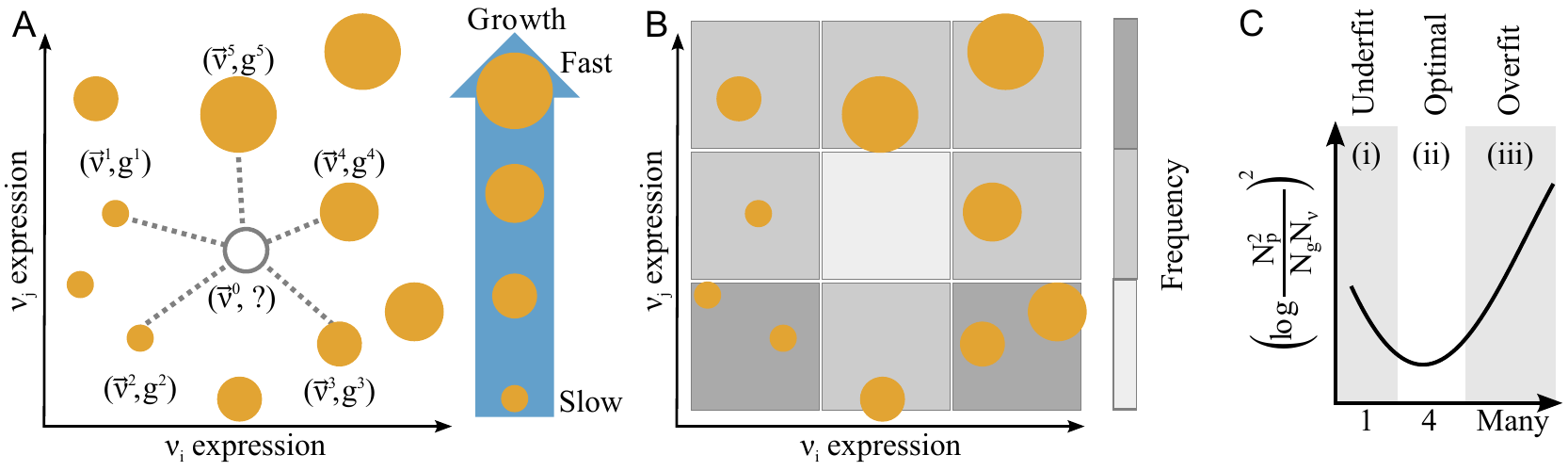}
\caption[Schematic illustrating growth rate estimation and eigengene selection.]{
Schematic illustrating growth rate estimation and eigengene selection. 
$(A)$~Measurements of gene expression and growth rate (location and size of solid circles, respectively) predict the growth rate of a hypothetical state (grey outlined circle) using the nearest measurements weighted by distance (dashed lines).
$(B)$~Discretized joint distributions of growth rate and eigengene (linear combinations of genes corresponding to the eigenvectors of the gene-gene correlation matrix) expression are indicated by size and grayscale background. The number of bins occupied ($N_{p}=8$), number of growth rate bins ($N_{g}=4$) and number of eigengene bins ($N_{\nu}=9$) characterize the efficiency.
$(C)$~ Sketch of the state-space occupancy plotted versus the number of eigengenes included in the model. Models with small $N_{\nu}$ lack enough information to differentiate between the $N_p$ observations. As more eigenvectors are added, $N_{\nu}$ increases geometrically, penalizing models using many eigengenes.
}
\label{5.1cartoon}
\end{center}
 %\vspace{-.5cm}
\end{figure*}

The method of KNN regression is a machine-learning technique trained on a set of paired measurements of independent and dependent variables that assigns an average of selected dependent variables to a test measurement of independent variables, where the dependent variables are selected by testing independent variables' similarity with the training measurements.
Here, the dependent variable is growth rate and the independent variables are the elements of a gene-expression profile as illustrated in \cref{5.1cartoon}A. The output of the KNN-fitting process is called a regressor, which takes gene expression as input and outputs an estimate of growth rate. From the set of all calculated eigengenes, we restrict to those with greatest potential to inform growth rate by searching for eigengenes that vary (possibly non-linearly) the most as growth rate changes. We discretize each measurement into bins of both growth rate and gene expression, and search for eigengenes that most evenly distribute the experiments into bins and thus span the range of variation observed.

\Cref{5.1cartoon}B illustrates in grids the joint distributions between growth rate and gene expression for two eigengenes. 
In this example, we suppose that each eigengene (and growth rate) can be in one of three states and discretize expression into these bins. 
Bins with higher densities of observations have darker colors.
Any single eigengene places limited constraints on the possible values of growth rate, but by adding more eigengenes 
the growth-rate possibilities for a given eigengene-expression profile narrow. The increased specificity comes at the cost of increasing the 
number of possible states an experiment could occupy, thereby increasing the sensitivity to noise (as diagrammed in \cref{5.1cartoon}C). 
Models that incorporate enough eigengenes to estimate growth rate but avoid overfitting maximize their ability to accurately explain the observed data and while retaining the ability to predict new data.

\subsection*{Optimization of KNN regressors}
%\vspace{-.2cm}
\label{dimred}

To select features that predict growth, we require an objective function that balances the explanatory and predictive capabilities of the regressor, called $G$. We first quantify the explanatory capabilities of $G$, which is characterized by the set of features $S$ that define the gene-expression subspace in which neighboring experiments are determined. The argument of $G^{(S)}$ is an experiment, $\vec{\nu}^{\,i}$, where as before $i$ is an index over experiments. We then determine the accuracy of $G^{(S)}(\vec{\nu}^{\,i})$ using the squared difference with the experimentally measured growth rate $g^i$.

As the number of eigengenes $|S|$ in $S$ increases, the predictions converge toward the measured growth rate, but the rate of convergence slows as models incorporate $|S| > 10$ eigengenes. Therefore, we introduce a criterion to quantify how efficient $G^{(S)}$ is in terms of state space. This term measures whether the decrease in error is large enough to justify the addition of another eigengene.
 In a  maximally efficient model,  each  unique combination of eigengene-expression levels  would have a corresponding  range of growth rates,  with no expression combination excluded.  
In other words, the number of bins occupied by the experiments in the dataset ($N_{p}$) would be equal to the number of growth rate bins in \cref{5.1cartoon}B ($N_{g}$). At the same time, $N_p$ would also be equal to the total possible number of configurations $N_{\nu}= \prod_1^{|S|} N_j$, where $N_j$ is the number of bins for the $j^{th}$ eigengene.  We take the square  of the (natural) logarithm of each ratio to obtain $\log [ N_p^2 / (N_g N_\nu) ]^2$, 
which we refer to as the state-space occupancy. Finally,  we introduce  the regularization parameter $\lambda$ to balance the relative contribution of the explanatory and predictive terms, yielding:
\begin{equation}
F(\vec{\nu},\vec{g}; |S|) = \underset{S}{\operatorname{\argmin}} \sqrt{\sum_i^E \left(G^{(S)}(\vec{\nu}^{\,i})-g^{i}\right)^2} + \lambda \left(\log \frac{ N^2_{p}}{N_{g} N_{\nu}} \right)^2. \label{optimization}
\end{equation}

 The value of $\lambda$ at the optimal value of $|S|$ is case-dependent and empirically found to be near $0.05$ for \emph{E. coli} (Methods and~Fig.~S1) and $0.001$ for \emph{S. cerevisiae}.
 Asymptotically, the optimal $|S|$ shifts toward smaller numbers as $\lambda \rightarrow \infty$ and  toward larger numbers as $\lambda \rightarrow 0$.

We compare the performance of  the various $G^{(S)}$  with Eq.~(\ref{optimization}) by dividing the dataset into training and test data consisting of gene-expression profiles paired with growth rate.
We choose to employ ``stratified, five-fold cross-validation''  which divides the existing data into subsets, called ``folds,'' whose distribution of growth rates is constrained to match the distribution of the entire dataset as closely as possible.
In testing the generalizability of the regressor, the dataset comprises five equally sized folds, and
four folds are used as training data to fit the regressor, which is tested on the fifth. 
Cross-validation is repeated with each fold used as test data once, yielding predictions for each of the $E$ experiments in the dataset. To account for variability in predictions due to fold construction, we average the predictions over 100 divisions of the dataset.

Finding the optimal set of eigengenes requires the testing of each possible set $S$, the number of which grows combinatorially. In view of the huge number of possibilities, we employ the ``forward-selection'' heuristic, which builds the set $S$ by adding eigengenes one at a time to find a set that is close to optimal~\cite{efron2004least}. Starting with $|S|=1$, and continuing for each size of $|S|$, we rank
all candidate regressors by the cross-validation procedure described in the previous paragraph. Next, we take the top-ranked set of eigengenes and form candidate sets of features of size $|S|+1$ by adding each of the remaining features to the top-ranked set of eigengenes. 
The cross-validation process continues until $|S|=50$, at which point the root mean squared error (RMSE) of the predictions has stopped improving.

\begin{figure}[t]
\begin{center}
\includegraphics[width=8.7cm]{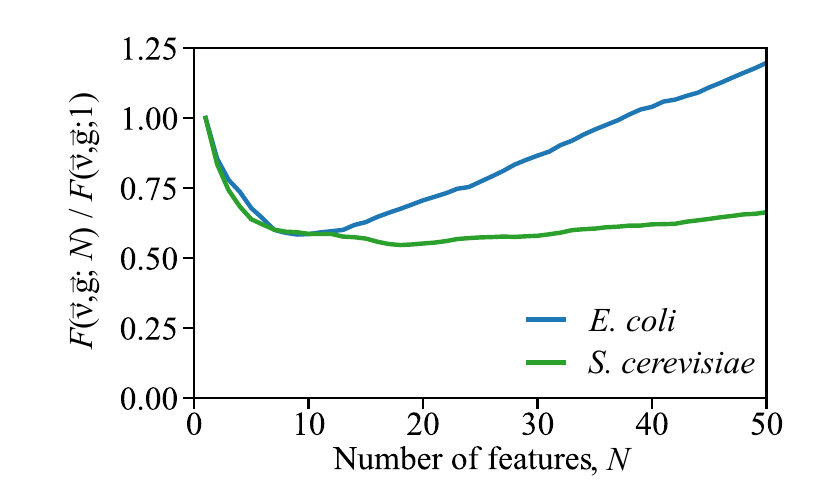}
\caption[Eigengene selection]{Selection of the best growth-predicting eigengenes in \emph{E. coli}  (blue,  $\lambda=0.05$)   and \emph{S. cerevisiae} (green, $\lambda=0.001$). {For presentation, values of $F(\vec{\nu},\vec{g}; N)$ (defined in Eq.~(\ref{optimization})) are scaled to their value at $N=1$.}}
\label{5.2dimred}
\end{center}
%\vspace{-.5cm}
\end{figure}

With the process for evaluating regressors in mind, we detail the optimal regressors for each value of $|S|$ in \cref{5.2dimred}. 
Strikingly, the models including 9 and  18   eigengenes out of thousands achieve a better balance of accuracy and predictability than by including much larger gene-based models in both \emph{E. coli} and \emph{S. cerevisiae} (\cref{tab:Precursortab}). 
 The small number of features at the minimum compresses the gene-expression information relevant to growth rate into a relatively low-dimensional subspace thereby facilitating further analysis.

For the best regressor composed of each number of eigengenes, we examine the trend in the square root of the sum of 
squared errors (SSE) (Fig.~S2A)) and the state-space occupancy (Fig.~S2B) corresponding to the first and second terms of Eq.~(\ref{optimization}), respectively. The SSE trend in Fig.~S2A shows the stagnating improvements in accuracy, despite the geometric decrease in the state-space occupancy by observations in Fig.~S2B. The large unoccupied fraction of state space hampers predictability as it is unclear how to extrapolate to hypothetical observations in this region.

\subsection*{Prediction comparison with existing methods}

In Table~\ref{tab:Precursortab}, we compare the quality of predictions based on eigengenes
with those based on the precursors of biomass---that is, all the genes included in the metabolic reconstruction
for each organism. We obtained the precursors of biomass from iJO1366~\cite{Orth2011}, containing 1,352 genes for \emph{E. coli},
and from Yeast~7~\cite{Aung2013}, containing 897 genes for \emph{S. cerevisiae}, and applied MI-POGUE to predict growth rate based on these genes' expression only.
For both organisms, models built on eigengenes have a higher coefficient of determination ($R^2$) and lower RMSE than those built on precursors of biomass, despite  requiring fewer features.

\begin{table}[b]
%\vspace{-.5cm}
\centering
\begin{threeparttable}[t]
\caption{Comparison of MI-POGUE's predictions when using optimized features versus precursors of biomass}
\begin{tabular}{p{3cm} p{4cm} p{1.5cm} p{1.5cm} r}
\toprule
Organism & Feature Set  & $R^2$ & RMSE & $|S|$  \\ 
\midrule
\emph{E. coli} & Optimized Features &  0.81\tnote{a} & 0.125 & 9 \\
                       & Biomass Precursors  & 0.76 & 0.138 & 1,352 \\ 
 \midrule
 \emph{S. cerevisiae} & Optimized Features  &  0.89\tnote{\,a}  & {0.028}  & {18} \\
                                   & Biomass Precursors  &  0.60  &  0.049   & 897 \\
\bottomrule
\end{tabular}
\begin{tablenotes}
    \item [ ] \emph{E. coli} data from ref.~\cite{Orth2011}. \emph{S. cerevisiae} data from ref.~\cite{Aung2013}.
    \item [a] Errors are smaller than $10^{-3}$.
\end{tablenotes}
\label{tab:Precursortab}
\end{threeparttable}
\end{table}

\begin{figure}[tb]
\begin{center}
\includegraphics[width=8.7cm]{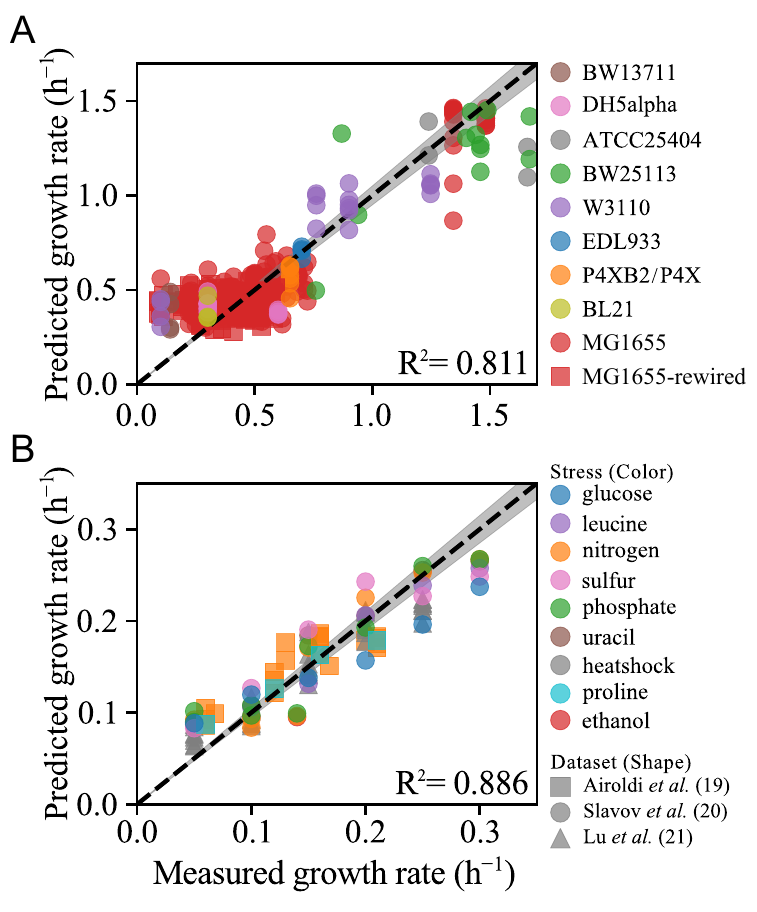}
\caption[Growth rate prediction accuracy]{Growth rate prediction accuracy in \emph{E. coli} and \emph{S. cerevisiae}. 
$(A)$~Scatter plot of predicted versus experimentally measured growth rates in \emph{E. coli} for the strains indicated in the legend  with the  5\% margin of measurement error indicated by gray shading. 
$(B)$~Same axes as in $(A)$, but for \emph{S. cerevisiae}, 
 with colors indicating the nutrient limitation or heat stress, and shapes indicating the dataset. 
For both organisms, $\lambda$ is the same as in \cref{5.2dimred}.}
\label{CGfig}
\end{center}
%\vspace{-0.5cm}
\end{figure}

Given the overall growth-rate prediction accuracy in  both organisms, we investigate the accuracy at the level of the individual experiments in \cref{CGfig}A. MI-POGUE performs comparably across most of the strains present in our dataset. 
 In \emph{E. coli}, 348 of 589 of the predicted experiments fall inside the 5\% error (grey region), despite systematic underestimation of the fastest and overestimation of the slowest growth rates. \emph{S. cerevisiae} shows a similar pattern, with a slightly higher fraction of experiments with less than 5\% error (66 of 107). The KNN approach systematically overestimates the slowest growth rates and underestimates the fastest growth rates because, by construction, the nearest neighbors of the slowest growth state will have faster-growth neighbors and vice-versa.  

Nevertheless, the level of accuracy motivated us to compare MI-POGUE with other existing methods using $R^2$ as a metric in Table~\ref{tab:Ecoli-comp-tab}.
MI-POGUE achieves agreement with experimental measurements superior to other reported methods without need for the additional step of converting the value of the biomass objective function to growth rate. MI-POGUE's accuracy outpaces that of CBMs 
 without requiring  metabolic uptake rates or enzyme kinetic parameters.

\begin{table}[tbhp]
\centering
\begin{threeparttable}[t]
\caption{Comparison of MI-POGUE's predictions with those of existing \emph{E. coli} methods}
\begin{tabular}{p{3cm} p{3cm} p{2cm} p{1cm}}
\hline
Method & $R^2$\,\tnote{a} & Reference  & $N$\tnote{b} \\
 \hline
 TRAME & \emph{0.36} & \cite{Carrera2014} & 24 \\
 ME-Model & \emph{0.25} &  &  \\
 iJO1366 & \emph{0.04} & &  \\
 MOMENT & \emph{0.49} &  & \\
 FBAwMC & \emph{0.13} &  &  \\
 \hline
 MOMENT & \emph{0.22} & \cite{Adadi2012}  & 24  \\
 FBAwMC & \emph{0.08} &    & 24  \\
 MOMENT & \emph{0.58} &   & 10  \\
 FBAwMC & \emph{0.64} &    & 10  \\
 \hline
RELATCH & \emph{0.48} & \cite{Kim2012} & 22  \\
 FBA & \emph{<0.01}\tnote{c} &  &   \\
 MOMA & \emph{0.17} &  &   \\
 ROOM & \emph{0.37}\tnote{c} &  &  \\
\hline
 MI-POGUE & 0.81  & This work &  589  \\
\hline
\end{tabular}
\begin{tablenotes}
   {\item [a] Italicized values quoted from reference.}
   {\item [b] Number of conditions tested.}
   {\item [c]  Correlation with measured growth is negative.}
\end{tablenotes}
\label{tab:Ecoli-comp-tab}
\end{threeparttable}
%\vspace{-.5cm}
\end{table}

We demonstrate the flexibility of our method by applying it to predict the  growth rate  of \emph{S. cerevisiae} grown in chemostats (\cref{CGfig}B). We also predicted growth using the linear models from ref.~\cite{Airoldi2009} for comparison  (SI~Methods). 
With $R^2 = 0.89$, MI-POGUE far outpaces the moderate success of the linear growth model for \emph{S. cerevisiae}~\cite{Airoldi2009}, which has $R^2 = 0.30$. 
Compared to \emph{E. coli}, MI-POGUE performs slightly better in \emph{S. cerevisiae}, despite the additional complexity of mapping gene expression to function~\cite{Ku2012}. This improvement in performance may be attributed to the relative sizes of the datasets and diversity of the conditions considered.  

 The success of MI-POGUE compared to the other methods derives, in part, from using eigengenes instead of single genes, but  at first one might think that this choice obscures the biological role of eigengenes. 
We note, however, that Weighted Gene Coexpression Network Analysis (WGCNA)~\cite{Langfelder2008} has recently been employed in fungi to associate gene modules with qualitative growth states~\cite{Baltussen2018}. 
Our approach is quantitative rather than qualitative but we can borrow this tool to interpret our eigengenes.
Specifically,  we develop a method to associate biological functions with eigengenes in \emph{E. coli} using WGCNA and Protein Analysis Through Evolutionary Relationships (PANTHER)~\cite{Mi2017}.  For each selected eigengene, we take the outer product of the eigengene with itself, resulting in a $J$ by $J$ similarity matrix that is rescaled such that the largest diagonal element is one (see Methods for full details). The rescaled matrix is subjected to WGCNA, yielding a module of genes associated with each eigenvector. We use PANTHER to identify the most over- and underrepresented Gene Ontology (GO) Biological Process annotations in each module (defined by $p < 0.05$, Fisher's Exact Test). 

We find that the top GO terms associated with modules derived from five of the nine eigengenes are overrepresented for polysaccharide, phospholipid, lipid, fatty acid, and amino acid metabolism. In two others, DNA repair and DNA metabolism are overrepresented. Of the remaining two, one is defined by its underrepresentation of metabolic genes, while the other has no terms meeting the $P$-value 
threshold---reflecting the sometimes weak association between eigengenes and GO terms.
That the nine selected eigengenes have largely nonoverlapping annotations is a result of the forward-selection process. Once an eigengene that captures one biological process is selected, it is less likely that a second eigengene capturing the same process will be selected.
The full gene lists and annotation lists are reported in Tables~\ref{data:geneLists}~and~\ref{data:interpretation}. 

%\vspace{-.3cm}
\section*{Discussion}
%\vspace{-.1cm}
MI-POGUE both addresses the challenges faced by previous methods and reduces the labor necessary to construct models that map gene expression to phenotype. It fully incorporates gene expression, relaxes the requirement that organisms be completely adapted to their environment, reduces reliance on metabolic uptake rates, and avoids the necessity of estimating enzyme kinetic parameters~\cite{Teusink2000,VanEunen2012,Chubukov2014}. Furthermore, MI-POGUE can be applied broadly, even in cases where the growth media are not strictly defined. Because MI-POGUE is flexible, it can repurpose previous measurements without requiring extensive targeted experiments to determine the structure of intracellular networks. Its ability to reduce the relevant features to a small number of eigengenes allows for genome-wide data to be expressed succinctly without loss of predictive power. These advantages are achieved while simultaneously improving the capacity to predict growth rate.

MI-POGUE can characterize the independence, synergy, or antagonism of perturbation pairs by evaluating the growth rate of a hypothetical transcriptional state constructed by adding the (experimentally measured) transcriptional responses of two perturbations to a reference state. Whereas local models leave open the possibility that some unobserved gene accounts for growth deviations from independence, our method implies that, within limitations of the available data, such deviations result from non-genetic mechanisms~\cite{PenalverBernabe2016}.

The approach of generating proposed states based on experimentally measured gene-expression responses to perturbations can also pre-screen experimental hypotheses. Such screening has the advantage of accounting for the real response of cells as opposed to the simulated response based on network structure~\cite{Alter2000,Brauer2008}. In the case of two perturbations where both are genetic, MI-POGUE can be used to estimate the growth of expression profiles resulting from both individual perturbations and the double perturbation, yielding a computational prediction of growth-rate epistasis, thereby providing a new tool to understand it~\cite{Segre2005a}. MI-POGUE could also be integrated with transcriptional regulatory network from databases~\cite{Gama-Castro2015} to predict the impact of previously unmeasured gene perturbations.

Researchers can incorporate MI-POGUE with existing strategies that interpret genetic networks in order to improve their effectiveness, as we demonstrate by using WGCNA and PANTHER to interpret the biological roles of eigengenes.
Genomic footprinting~\cite{Gerdes2003}, previously used to find essential genes, could be used to resolve regions of gene expression that yield zero growth, which can enhance the ability of the KNN algorithm to extrapolate beyond the training data.

The applicability of MI-POGUE to metabolic engineering, antibiotic development, and systems biology 
is expected to encourage its adoption and further refinement or the adoption of similar methods. For example, metabolic engineers could tailor MI-POGUE to offer predictions of a key uptake or secretion rate based on the organism's gene expression. Antibiotic developers could use transcriptional changes in response to drugs to choose combinations that result in the slowest growth rate as predicted by MI-POGUE.  Systems biologists could use MI-POGUE to look for interactions between genes by taking transcriptional responses to single knockouts, adding them, and simulating the outcome. 
We also note that the mapping of eigengenes to biological functions as described here merit further investigation. Using modern community-detection algorithms, such as weighted stochastic block models~\cite{Aicher2014}, can help discern finer-scale structure of gene modules

Given that cells are complex systems, weak and indirect interactions at the molecular level can influence behavior at the whole-cell level. 
Reductionist strategies are poorly suited to study these phenomena, but approaches like MI-POGUE, which combine machine learning with bioinformatic ``big data,'' have the potential to capture these subtle effects. As systems biologists adapt machine-learning techniques to better interpret high-throughput data, the new interpretative power of these techniques has the potential to reveal under-appreciated and sometimes counter-intuitive effects that will drive the field into the future.
%\vspace{-.1cm}
\section*{Methods}
%\vspace{-.5cm}
\subsection*{Implementation of  MI-POGUE}
\label{correlations}

We used an implementation of the nearest-neighbors algorithm~\cite{Altman1992} found in the Python 
sklearn package~\cite{Pedregosa2011}.
For \emph{E. coli} (\emph{S. cerevisiae}) models, we chose $k=7$ ($k=8$)  to be the number of neighbors as this number was shown to perform better than other choices (SI~Appendix, Figs.~S3~and~S4). Additional details about the development and extensions of MI-POGUE are described in the SI~Appendix, with the code and instructions for running MI-POGUE is available in ref.~\cite{Wytock2018}. 

%\vspace{-.1cm}
\subsection*{Discretization of data}
The growth-rate bins are fixed so that the number of experiments in each bin is 
approximately the same. In addition, for a given feature, every tenth percentile (that is, the $10^{th}, 20^{th}, \dots, 90^{th}$) of the projection of gene expression onto that feature
is calculated from the available data. If the difference between consecutive percentiles (that is, the bin width) is larger than 10\% 
of the mean of the consecutive percentiles (the bin midpoint), then the bin is left in place. 
When the width is smaller, we randomly choose either the previous or subsequent percentile and merge 
the data into a larger bin, recalculating the width and midpoint. The bin-merging procedure continues until all bins' widths
are larger than 10\% of their midpoints. 

%\vspace{-.15cm}
 \subsection*{Precursors of biomass} 
We downloaded the metabolic model iJO1366~\cite{Orth2011} from \emph{E. coli} and all those from \emph{S. cerevisiae} considered in ref.~\cite{Heavner2015} from the supplementary material provided with the associated publications. 
%\vspace{-.15cm}

\subsection*{Experimental data} 
For \emph{E. coli}, we downloaded gene-expression data and experimental metadata from ref.~\cite{Carrera2014}.
A lightly edited version of supplementary~table~2 from ref.~\cite{Carrera2014} describing the full \emph{E. coli} dataset can be found at the author's GitHub.

For \emph{S. cerevisiae}, we downloaded gene-expression data and growth-rate data from ref.~\cite{Airoldi2009}, packaged as an ``.RData'' archive in the dataset~S1 in that reference. Loading the archive into R, we used the data frames that reported gene-expression data for strains growing at a fixed rate: ``frmeDataCharles''~\cite{Lu2009} and ``frmeDataGresham''~\cite{Airoldi2016}. To these, we added data from ref.~\cite{Slavov2011} (downloaded from \url{http://genomics-pubs.princeton.edu/grr/}), whose growth-rate, but not gene-expression, data are included in the R archive. These three datasets shared 5,527 unique genes and 107 total experiments.

For the purpose of estimating the correlations between genes in \emph{S. cerevisiae}, we obtained data from two large-scale screens of gene knockouts~\cite{Hughes2000,Kemmeren2014}. The 300 expression profiles of~\cite{Hughes2000} were used as provided in the RData archive. 
Raw data from ref.~\cite{Kemmeren2014} were downloaded from GEO and preprocessed as described in the supplement of that reference. 
 Following ref.~\cite{Kemmeren2014}, we excluded gene-expression profiles with fewer than four genes with significant responses to the gene deletion. These 1,369 experiments comprising 700 responsive strains were identified from supplemental~table~S1 of ref.~\cite{Kemmeren2014}. 
%\vspace{-.2cm}
\subsection*{Determining optimal parameter values}
The optimal number of neighbors was empirically determined in two ways: first by starting at the previously identified set of optimal eigengenes for $k=7$ and cross-validating the dataset with different numbers of neighbors, and second by re-running feature selection with an optimal number of neighbors obtained from the first method.  The first case is illustrated by SI~Appendix, Fig.~S3 for a range of values for $k$, the number of nearest neighbors. We repeated the feature selection for $k=5$, determined the maximum for $R^2$ of the peak, and found that this performs less well than the $k=7$ case.

We repeated the feature selection for various values of the regularization parameter $\lambda$, which controls the relative weighting of the prediction error and the state-space occupancy terms of Eq.~(\ref{optimization}) as shown in SI~Appendix, Fig.~S1. Since the forward-selection algorithm finds the best-fitting eigengene at each stage, changes to $\lambda$ and the fold divisions can lead to different selections from the available eigengenes. As before, the variability in eigengenes due to fold selection can be mitigated by repeating the selection and taking the eigenvector that performs best on average. In \emph{E. coli}, the maximal $R^2$ achieved for $\lambda = 10^{-2}$ in both the $k$-fold and leave-one-GSE-out cross-validation strategies is greater than that achieved by the other values. Additional tests led to us selecting $\lambda=0.05$. A similar approach was used to select $\lambda$ for the \emph{S. cerevisiae} dataset.

%%\vspace{-.2cm}
%\showmatmethods 

\section*{Acknowledgements}
This work was supported by NIH/NIGMS R01GM113238 and NIH/NCI 1U54CA193419. 
TPW also acknowledges support from NSF-GRFP fund No. DGE-0824162 as well as 
NIH/NIGMS 5T32GM008382.
%%\vspace{-.2cm}
%\showacknow 
%%\vspace{-.3cm}
%\pnasbreak

\section*{Author contributions}
T.P.W. and A.E.M. designed research; T.P.W. performed research;
T.P.W. and A.E.M. analyzed data; and T.P.W. and A.E.M. wrote the paper.

\bibliographystyle{pnas-new}
\bibliography{msb_working_paper}
\newpage

\newpage
\pagebreak[4]
\section*{Supplemental Information}

\setcounter{figure}{0}
\setcounter{table}{0}
\setcounter{equation}{0}
\renewcommand{\thesection}{S\arabic{section}}   
\renewcommand{\thetable}{S\arabic{table}}   
\renewcommand{\thefigure}{S\arabic{figure}}
\renewcommand{\theequation}{S\arabic{equation}}

\subsection*{Overview}
The SI Results are an extended description of the parameter estimation, cross-validation, and eigengene interpretation that we performed on MI-POGUE. The SI Methods describe the implementation of the linear models, present additional context for MI-POGUE's objective function, and provide guidance on how to select the parameter $\lambda$.

\section*{SI Results}

\subsection*{Alternative cross-validation strategies}
In the main text, we adopt a stratified $k$-fold cross-validation strategy. This strategy can be thought of an upper bound of MI-POGUE's accuracy. In \emph{E. coli}, we additionally investigate the possibility of overfitting using a strategy we call ``leave one GSE out'' in which all the gene-expression and growth-rate measurements associated with a particular Gene Expression Omnibus~\cite{Barrett2012} (GEO) Series accession number (i.e., GSE) are withheld from the training set (see black dashed curves reproduced in \cref{supp_fig:var-lambda,supp_fig:opt-v-knn,supp_fig:opt-v-noise,supp_fig:corr-v-gene}). In \emph{S. cerevisiae} the equivalent method is ``leave one group out.'' This strategy is the most stringent and is equivalent to applying MI-POGUE to unseen data. In this case, the peak values for $R^2$ are in the range of 0.45 for \emph{E. coli} and 0.36 for \emph{S. cerevisiae} (green curve in \cref{supp_fig:yeast-cv}). 

Specifically in \emph{S. cerevisiae}, we implement two more strategies. In the first, we excluded the set of experiments with a particular  growth rate (orange curve in \cref{supp_fig:yeast-cv}). In the second, we excluded all experiments undergoing a particular treatment; for example, all strains grown in phosphate limiting conditions (red curve in \cref{supp_fig:yeast-cv}). These two cases exhibit different behaviors as the number of features used in the regressor increases. The predictions seem to improve slightly in the case of an excluded growth rate, because the additional features aid interpolation of the growth rate. Conversely, caution is needed when extrapolating to new treatments, as adding more features in this case causes accuracy to decline due to overfitting.

\subsection*{Using genes instead of eigengenes}
We sought to establish whether models formed with eigengenes performed better than those formed with genes by performing feature selection in both instances and comparing the resulting models. The results are shown in \cref{supp_fig:corr-v-gene}. In the stratified $k$-fold case, gene-based models appear to outperform eigengene-based models for models with less than 10 features. However, eigenegene-based models remain preferable because they achieve a higher $R^2$ at large feature numbers in the stratified $k$-fold case, they achieve a higher $R^2$ in the leave-one-GSE-out case, and they have a smaller number of features to search through, which reduces the computational time. We note that the selected eigengenes are not individually correlated with growth (Table~\ref{tab:EigenvalueRanks}). This is a reflection of the non-linear and non-parametric nature of KNN regression.

\subsection*{Noise sensitivity}
The optimal features include eigenvectors associated with small eigenvalues. These small eigenvalues tend to be sensitive to the level of noise included in the features. The effect of noise can be simulated by first calculating the mean and variance of each gene across the 2,196 experiments. This mean and variance are used to define a Gaussian distribution. We use this distribution to generate ``pseudo-profiles'' (i.e., simulated data) and model the effect of noise by including four pseudo-profiles per experiment in the training set when testing the KNN regressor. 
Each pseudo-profile is assigned a growth rate that is generated by taking the actual measurement as the mean and imposing a 5\% error rate about this mean. The pseudo-profiles are then projected onto the previously calculated eigengenes. The effect of noise on the eigengenes selected is illustrated in \cref{supp_fig:opt-v-noise}. As expected, the inclusion of noise shifts the eigenvalues associated with the selected eigengenes toward those that are larger in magnitude. In addition, the ability to predict growth suffers, both as more features are added in the stratified $k$-fold case (\cref{supp_fig:opt-v-noise}A), and especially in the leave-one-GSE-out case (\cref{supp_fig:opt-v-noise}B).

\subsection*{Sensitivity of eigenvector selection}
It is important to note that the eigenvectors selected to predict growth rate are not a unique set. The choice to include noise or use a different value for $\lambda$ lead to different eigenvectors being chosen. Furthermore, adding or subtracting experiments from the data used to calculate correlations necessarily changes the eigenvectors and eigenvalues. 

Because the forward-selection algorithm finds the best-fitting eigenvector at each stage, different 
breakdowns of the cross-validation can change the selected eigenvector. In the version of MI-POGUE that incorporates the role of noise, fluctuations in the pseudo-profiles can likewise change the identity of the best-fitting eigenvector. Therefore, 
selection of the best feature must be repeated multiple times to account for the variability. Efforts to account for the uncertainty associated with each datapoint yield sets of features that perform less well in the leave-one-GSE-out case (see \cref{supp_fig:opt-v-noise}), underscoring the challenge of extending the model to predict outside data. Some of this prediction error could be mitigated with improved sampling of various stressful states.

As the predictions of growth rate are robust to the eigenvectors chosen, it appears that the eigenvectors of the gene-gene correlation matrix are related only indirectly to the biological underpinnings of the gene regulatory network. Currently, it is unclear whether there are better ways to decompose the gene expression that still produce accurate estimates of growth rate in a low-dimensional space.
Furthermore, it is uncertain that more faithfully reproducing biological details will result in decompositions of gene expression that more accurately predict growth rate. The former uncertainty is well suited for additional study in machine learning, and a potential application for deep neural networks, assuming that enough data is available. The latter problem of developing biologically faithful models that still predict growth rate is one for systems biologists to systematically incorporate other sources of bioinformatic data, including the results of CBMs to improve the prediction of growth rate. In particular, incorporating other biological data sources will enable \emph{in silico} prediction of the effects of genetic and environmental perturbations to the system. Nevertheless, we take a first step toward linking the eigenvectors to biological pathways in the next section. 

\subsection*{Interpretation of eigengenes}
 We adapt Weighted Gene Coexpression Network Analysis (WGCNA)~\cite{Langfelder2008} toward the interpretation of eigenvalues. Our strategy is to create a similarity measure based on each eigenvector, apply WGCNA on this measure, and then apply an annotation analysis method (PANTHER) on the resulting modules.
 
 Each eigenvector, $\mathbf{p}_l = \mathbf{P}^{l}$, can be transformed into a similarity measure using
 the outer product $\mathbf{A}^l =  \mathbf{p}_l \mathbf{p}_l^{\intercal}$ and scaling by the reciprocal of the largest diagonal element $a_l = 1/\max_i{A^l_{ii}}$, yielding
 \begin{equation}
  \mathbf{S}^{l} =\mathbf{A}^l a_l. \label{eq:similarity}
 \end{equation}
The similarity is a rescaled version of the eigenvector's independent contribution to the overall correlation matrix. The rescaling is necessary to ensure that the application of WGCNA results in a connected network.

WGCNA applies soft thresholding by applying $|s_{ij}|^\beta$, where the $s_{ij}$ are the elements of Eq.~(\ref{eq:similarity}), and $\beta=5$ is chosen such that the weighted degree distribution follows a power law. The choice of using absolute value corresponds to the ``unsigned'' option for constructing the weighted adjacency matrix from which the topological overlap matrix~\cite{Ravasz2002}, $T$, is calculated. Applying hierarchical clustering to $1-T$ using the ``average'' linkage method, we find modules—sets of genes that are more connected to one another than to the rest of the network (Table~\ref{data:geneLists}). The genes from each module are subjected to PANTHER~\cite{Mi2017} (accessed at \url{http://pantherdb.org/tools/uploadFiles.jsp}) to find overrepresented annotations in the modules, which hints at the function of the eigengene. For each selected eigenvector, overrepresented annotations are reported in~Table~\ref{data:interpretation}.

\section*{SI Methods} 

\subsection*{Linear models}
The supplemental~dataset~S1 of ref.~\cite{Airoldi2009} includes a function ``calculateRates'' in the namespace that estimates the growth rate from a gene-expression dataset based on a linear model of gene expression. The arguments of calculateRates are a gene-expression dataset, growth-rate parameters, and a calibration list of genes. The gene-expression datasets are the data frames obtained as described in the ``Experimental Data'' section of the Methods. The supplemental archive in ref.~\cite{Airoldi2009} includes ``frmeGRParameters'' and ``lsCalibration,'' which supply the growth-rate parameters and calibration list, respectively. We call calculateRates on each of the three datasets to obtain the predicted growth rates. The real growth rates are included in the supplemental archive as ``vdRealCharles,'' ``vdRealGresham,'' and``vdRealSlavov.'' We take the square of the correlation coefficient of the real rates with the calculated rates to get the value of $R^2$.

\subsection*{Motivation for the objective function}
We adapt a typical method for enforcing sparsity, known as Tikhonov regularization or ridge regression in the statistics 
community~\cite{golub1999tikhonov}, which imposes an $\ell_2$ regularization to select among solutions of an ill-posed least squares 
problem. In the context of linear least squares problems, Tikhonov regularization
adds a term similar to the second Eq.~(\ref{optimization}) consisting of the squared magnitude of the linear coefficients to the least squares
term (the first in Eq.~(\ref{optimization})). In contrast to linear regression, KNN regression has no parameters to fit. 
In place of these parameters, we focus on the state-space occupancy of the joint distribution of discretized gene expression and growth rate (see Fig.~1B,C). 

Therefore, we arrived at Eq.~(\ref{optimization}), which introduces parameters fixed by the organism ($M$), 
the dataset ($E$, $N_{g}$, $N_{\nu}$, $g^{i}$) or a combination of the two ($\vec{\nu}^{\,i}$, $N_{p}$), in addition to a parameter $\lambda$.
The set, $M$, is the set of all eigengenes. It is a pool from which subsets $S$ of 
fixed size $|S|$ are chosen, and the pool of genes whose expression composes the eigengenes are defined by those present in the transcriptomics chip. 
Each organism's dataset has a fixed number of experiments, $E$, 
and the bins of growth rate $N_{g}$ are determined by the accuracy of the measurement and the distribution
of sampled growth rates, $g^{i}$. Likewise, the measurement error and observed distribution fix the number 
of gene-expression bins, $N_j$. 
Among the total number of state space bins, $N_{g}\prod_j^J N_j$, 
$N_{p}$ are occupied by at least one experimental observation. The specific choices of the eigengenes 
to include in $S$ will determine the discretization of the gene-expression space, the fraction of the state space
containing at least one observation, and the agreement of the growth-rate prediction $G^{(S)}(\vec{\nu}^{\,i})$ 
for each experiment $\vec{\nu}^{\,i}$ with the measured growth rate $g^{i}$. 

\subsection*{Choosing the parameter $\lambda$}
In the Results, we briefly describe the limiting behavior for $\lambda$, which is chosen to balance the least squares term (first) with 
the state-space occupancy term (second). Here, we go into further detail regarding how to empirically select $\lambda$. We first note that as $\lambda \rightarrow \infty$, the eigengenes that have a coarse-grained one-to-one correspondence (see Fig.~1B of the main text) with growth rate are selected, but as $\lambda \rightarrow 0$ the accuracy becomes the determinative factor. Therefore, there exist constants $A$ and $B$ such that for values of $\lambda > A$ the order of selected eigengenes is the same, and likewise for  $\lambda < B$.

\begin{figure}[tb]
\begin{center}
\includegraphics[width=114mm]{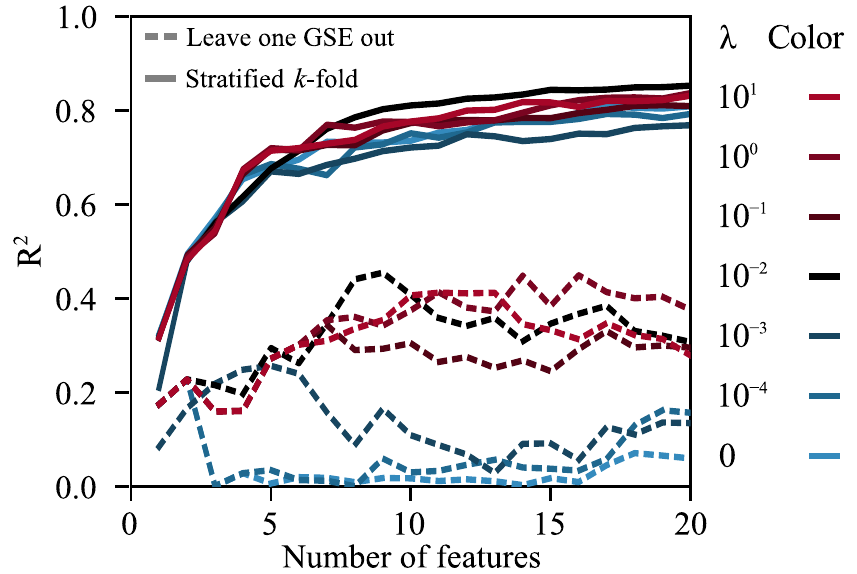}
\caption{Coefficient of determination ($R^2$) between predicted and measured growth rate  in \emph{E. coli} plotted as a function of the number eigengenes for various values of $\lambda$ as indicated by color. Line styles denote the cross-validation strategy.  Solid lines signify stratified $k$-fold cross-validation described in the text, while dashed lines signify the ``leave-one-GSE-out'' strategy in which all the gene-expression and growth-rate measurements associated with a particular GEO Series accession number (i.e., GSE) are withheld from the training set. The model is built on the remaining measurements. The withheld expression measurements are used to predict the withheld growth rates. Data associated from each GSE is withheld once.  }
\label{supp_fig:var-lambda}
\end{center}
\end{figure}

Starting from Eq.~(\ref{optimization}) of the main text, we can bound constants $A$ and $B$ as follows in terms of $N_g$, $N_j$, $|S|$, $E$, and the expected accuracy of the growth-rate estimation, $\alpha$. For simplicity, we approximate $N_j$ as a constant, as we observe in our datasets. We can rewrite the first term of Eq.~(\ref{optimization}) as $\alpha E$. Next, we solve for the value of $\lambda$ for which the two terms are equal if $|S|=1$ and if $|S|  \gg 1$ which result in the lower bound for $A$ and upper bound for $B$, respectively. The first result is that $ A > (\alpha E)/\left(\log \left(\frac{E^2}{N_g N_j} \right) \right)^2$, and the second is that $B < (\alpha E)/(|S| \log N_j)^2$. Plugging in the values for the \emph{E. coli} dataset ($E = 589$, $N_g = 16$, $N_j = 10$), and assuming $A > 1$ in the first case and, letting $|S|=50$, $B <  0.005$. The range tested in Fig.~\ref{supp_fig:var-lambda} extends slightly beyond [0.005,1], with the best fit value lying near the geometric mean.

\newpage

\pagebreak
\newpage

\begin{figure}[t]
\begin{center}
\includegraphics[width=114mm]{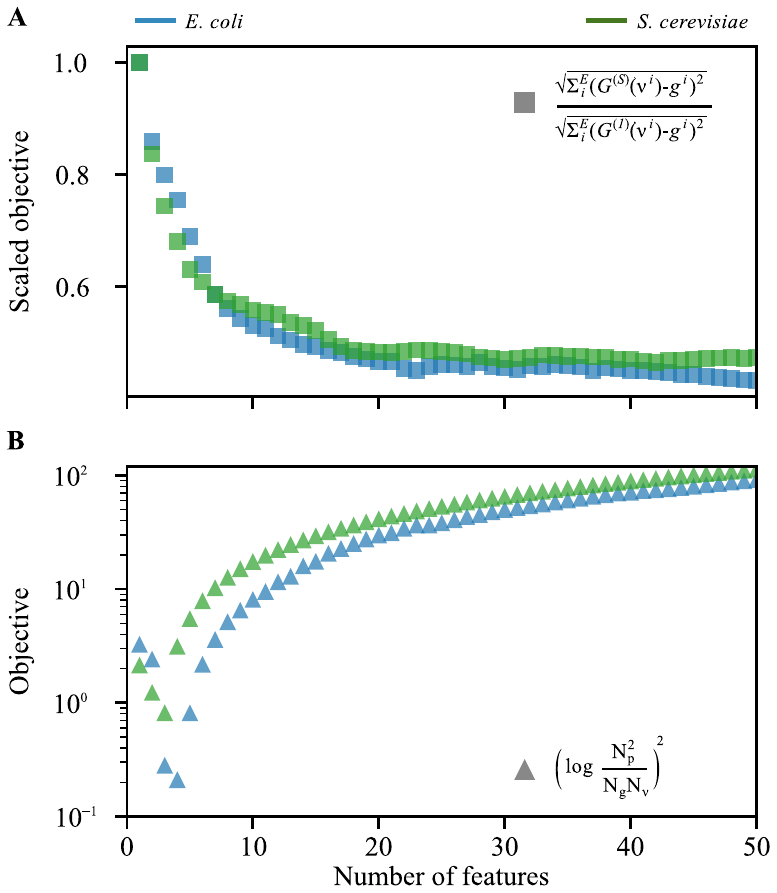}
\caption{Breakdown by term of the eigengene models in Fig.~2. 
The flattening of the squared-error term $(A)$ with increasing numbers of features, coupled with the increase in the state-space occupancy term $(B)$, leads to optimal models for small numbers of eigengenes, especially as $\lambda \rightarrow \infty$. In $A$, the square-error term is scaled to the value achieved for a single feature to facilitate presentation.}
\label{supp_fig:5.2breakdown}
\end{center}
\end{figure}

\pagebreak
\newpage

\begin{figure}[tb]
\begin{center}
\includegraphics[width=114mm]{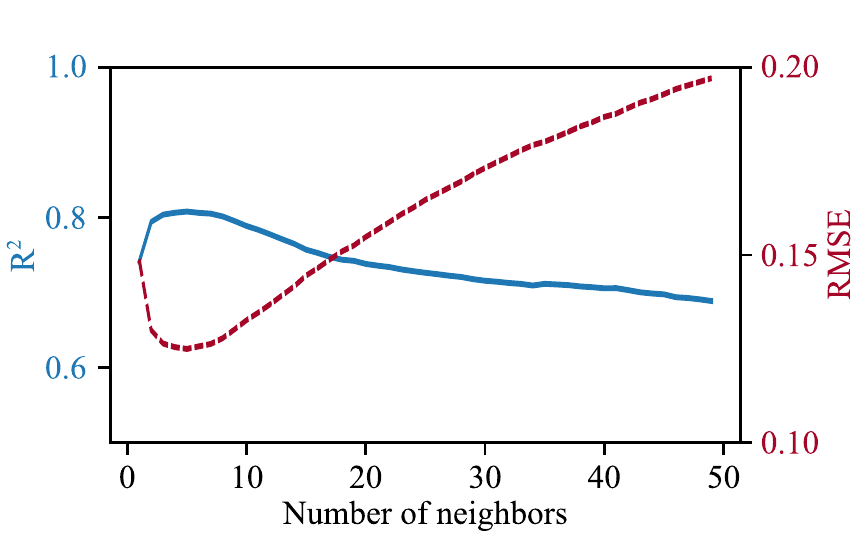}
\caption[Parameter sweeps in \emph{E. coli}]{Coefficient of determination (left axis, blue) and RMSE (right axis, red) as a function of the number of neighbors in \emph{E. coli} for models with 9 features. The chosen value of $k=7$ is near the maximum.}
\label{supp_fig:neighbors-ecoli}
\end{center}
\end{figure}

\pagebreak
\newpage

\begin{figure}[tb]
\begin{center}
\includegraphics[width=114mm]{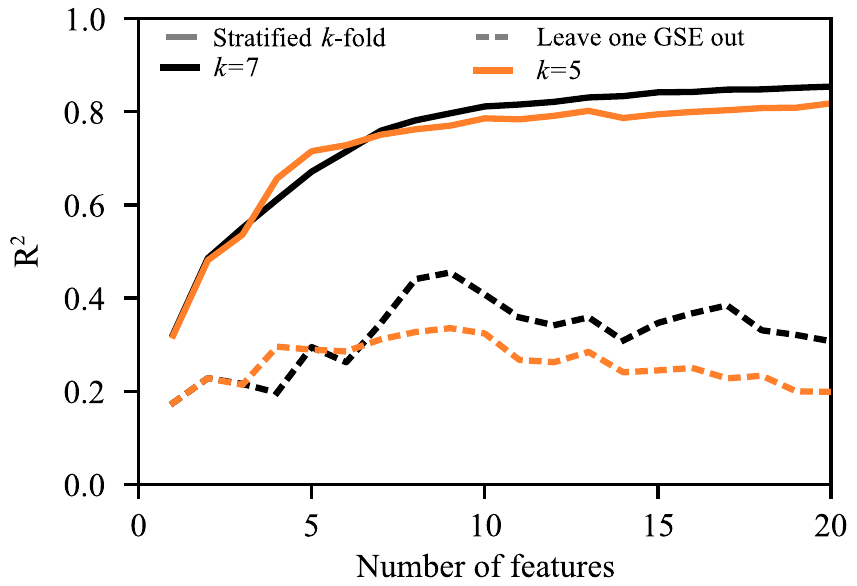}
\caption[Comparsion of $k=5$ versus $k=7$]{Coefficient of determination ($R^2$) as a function of the number of features in \emph{E. coli} for models constructed with $k=7$ (black) and $k=5$ (orange). Although $k=5$ appears to be the best performing value for this parameter in \cref{supp_fig:neighbors-ecoli}, feature selection conducted with $k=7$ leads to more accurate regressors, as demonstrated here. Line styles denote the cross-validation strategy as described in the caption of \cref{supp_fig:var-lambda}.
  }
\label{supp_fig:opt-v-knn}
\end{center}
\end{figure}

\pagebreak
\newpage

\begin{figure}[tb]
\begin{center}
\includegraphics[width=114mm]{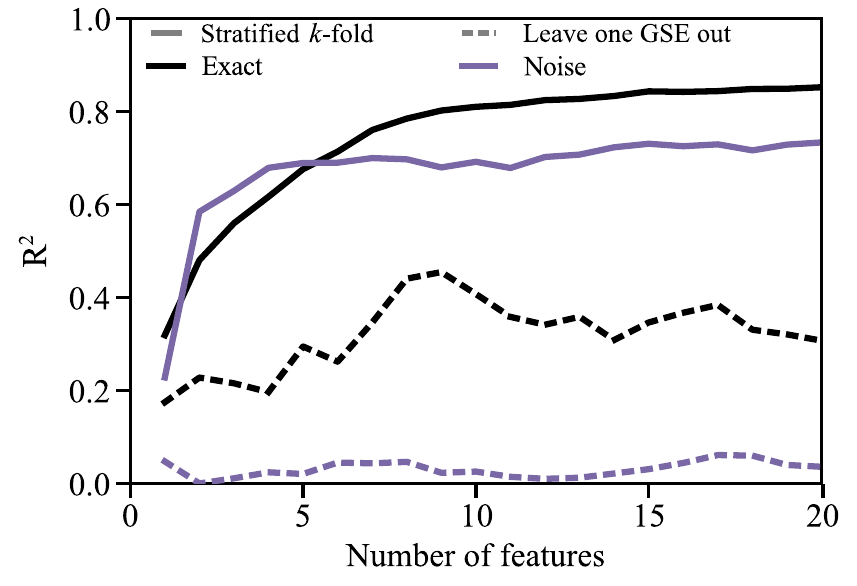}
\caption[Comparison of genes and eigengenes]{Coefficient of determination ($R^2$) between predicted and measured growth rate  in \emph{E. coli}, plotted as a function of the number 
of eigengenes when the feature selection is run with (purple) or without (black) noise.
Line styles denote the cross-validation strategy as described in the caption of  \cref{supp_fig:var-lambda}.
}

\label{supp_fig:opt-v-noise}
\end{center}
\end{figure}

\pagebreak
\newpage

\begin{figure}[tb]
\begin{center}
\includegraphics[width=114mm]{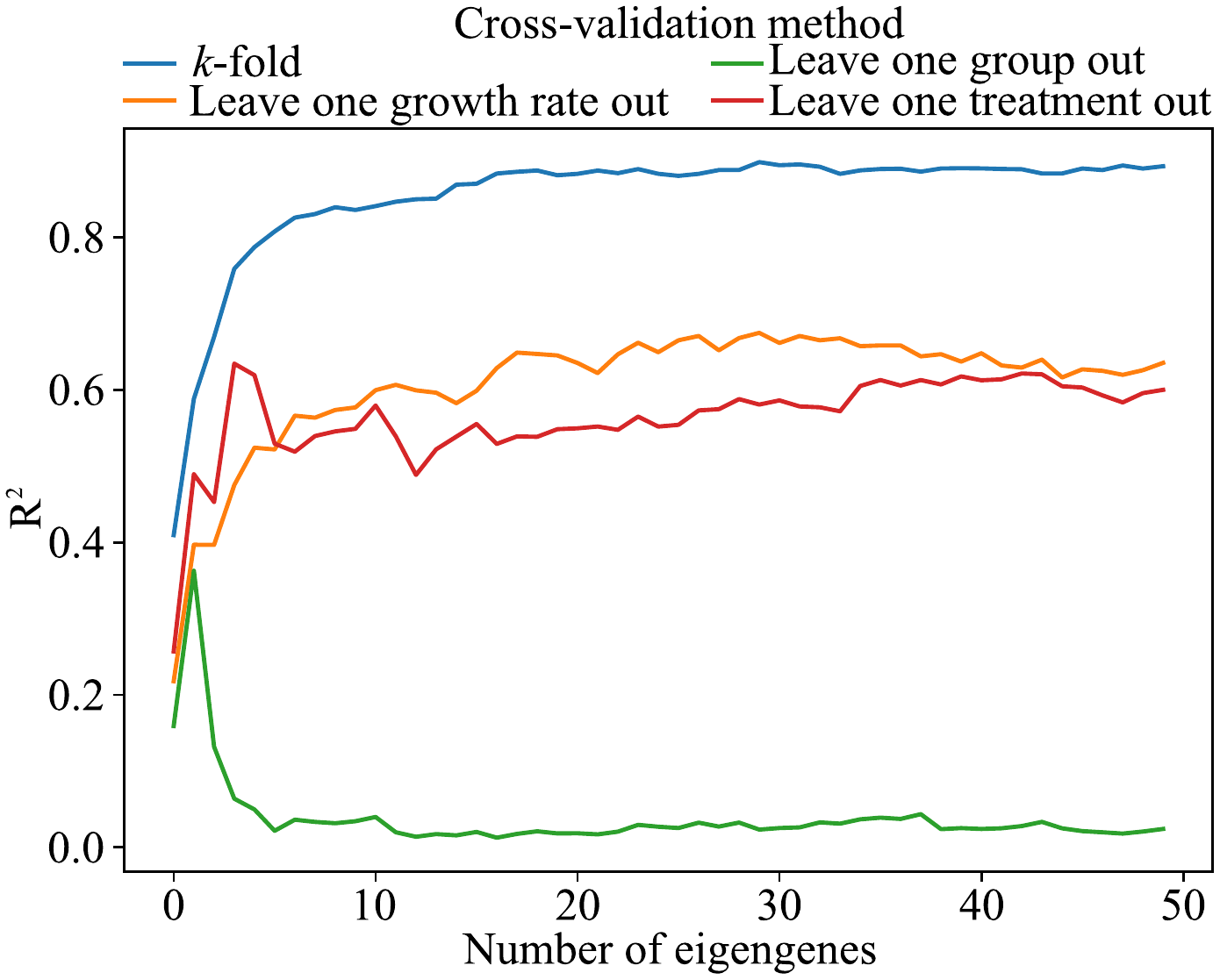}
\caption[\emph{S. cerevisiae} cross-validation]{
Coefficient of determination ($R^2$) between predicted and measured growth rate  in \emph{S. cerevisiae}, plotted as a function of the number of eigengenes. The color of the line indicates the cross-validation strategy used. In the ``$k$-fold'' strategy, MI-POGUE is trained on a randomly chosen four-fifths of the data and tested on the remaining fifth. The ``leave-one-growth-rate/-treatment/-group-out'' strategy is analogous to the ``leave-one-GSE-out'' strategy employed in \emph{E. coli} (described in the caption of \cref{supp_fig:var-lambda}) in that the test set comprises all experiments with a particular growth rate/treatment/dataset and MI-POGUE is trained on the remaining data. The declines in $R^2$  for the ``leave-one-group-out'' and the ``leave-one-treatment-out'' for larger numbers of features are the result of overfitting.
}
\label{supp_fig:yeast-cv}
\end{center}
\end{figure}

\pagebreak

\begin{figure}[tb]
\begin{center}
\includegraphics[width=114mm]{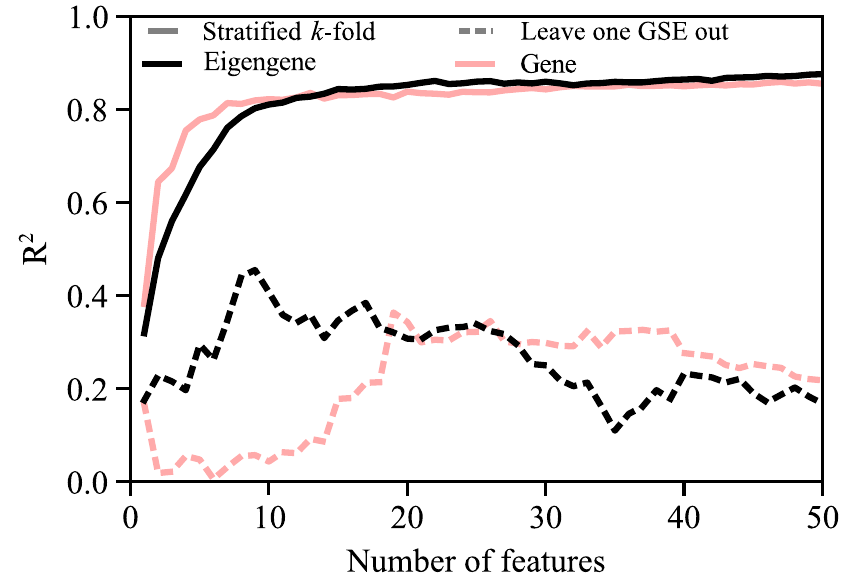}
\caption[Comparison of genes and eigengenes]{Coefficient of determination ($R^2$) between predicted and measured growth rate  in \emph{E. coli}, plotted as a function of the number of either genes (pink) or eigengenes (black) used. Line styles denote the cross-validation strategy as described in the caption of \cref{supp_fig:var-lambda}.  }
\label{supp_fig:corr-v-gene}
\end{center}
\end{figure}

\FloatBarrier

%\dataset{MI\_POGUE\_CODE.tar.gz}{\label{data:source}Source code, data, and scripts containing scripts to run the analyses in the paper are available at \url{https://github.com/twytock/MI-POGUE}. The scripts are designed to download code from the references as prescribed in the Methods.}

%\dataset{Analysis\_recipe.pdf}{\label{data:recipe}Instructions for running the scripts in MI\_POGUE\_CODE.tar.gz.}
\begin{table}[]
\caption{Correlation between the selected eigengenes in \emph{E. coli} and growth rate. \rm{The columns are: the eigenvalue of the selected eigengene, the corresponding rank by the absolute value of the Spearman correlation between selected eigengenes and growth rate (among all eigengenes), and the value of correlation coefficient. Lower rank numbers correspond to larger correlation (in absolute value). ``Including Noise'' lists the eigenvalues selected when noise is considered during the feature selection.}} \label{tab:EigenvalueRanks}
\begin{tabular}{lrr}
\toprule
Selected Eigenvalue  & Rank & Spearman Correlation \\
\midrule
0.053516       & 1829 & -0.040         \\
0.052508       & 990  & 0.157          \\
0.060603       & 310  & -0.264         \\
0.060767       & 2125 & 0.002          \\
0.159468       & 1994 & -0.019         \\
0.077524       & 1287 & -0.113         \\
0.095336       & 1355 & 0.105           \\
0.098494       & 768  & -0.188         \\
0.079097       & 542  & -0.222         \\
\midrule
(Including Noise)   &      &  \\
\midrule
32.221929      & 50   & 0.354          \\
\ \ 3.039641       & 1896 & -0.032         \\
\ \ 3.114244       & 1519 & 0.083          \\
\ \ 1.488373       & 1081 & -0.143         \\
\ \ 3.386905       & 1963 & 0.023  \\ 
\bottomrule   
\end{tabular}
\end{table}

\begin{table}
\caption{\label{data:geneLists} {\bf TableS2.xlsx} Lists of genes in the modules found for each selected eigengene. Eigengenes are labelled by their eigenvalue and presented in the order of their selection (i.e., Column B is the first selected eigengene). Genes in each list are identified by their gene symbol.}
\end{table}

\begin{table}
\caption{\label{data:interpretation} {\bf TableS3.xlsx} The results of applying PANTHER to the gene lists. Each table of biological processes is in a separate spreadsheet. The ``Legend'' spreadsheet describes the meaning of the columns.}
\end{table}
\end{document}